\documentclass[twoside,twocolumn,9pt]{article}
\usepackage{extsizes}
\usepackage[super,sort&compress,comma]{natbib} 
\usepackage[version=3]{mhchem}
\usepackage[left=1.5cm, right=1.5cm, top=1.785cm, bottom=2.0cm]{geometry}
\usepackage{balance}
\usepackage{mathptmx}
\usepackage{sectsty}
\usepackage{graphicx} 
\usepackage{lastpage}
\usepackage[format=plain,justification=justified,singlelinecheck=false,font={stretch=1.125,small,sf},labelfont=bf,labelsep=space]{caption}
\usepackage{float}
\usepackage{fancyhdr}
\usepackage{fnpos}
\usepackage[english]{babel}
\addto{\captionsenglish}{%
  
}
\usepackage{array}
\usepackage{droidsans}
\usepackage{charter}
\usepackage[T1]{fontenc}
\usepackage[usenames,dvipsnames]{xcolor}
\usepackage{setspace}
\usepackage[compact]{titlesec}
\usepackage{hyperref}

\definecolor{cream}{RGB}{222,217,201}

\begin{document}

\pagestyle{fancy}
\thispagestyle{plain}
\fancypagestyle{plain}{
\renewcommand{\headrulewidth}{0pt}
}

\makeFNbottom
\makeatletter
\renewcommand\LARGE{\@setfontsize\LARGE{15pt}{17}}
\renewcommand\Large{\@setfontsize\Large{12pt}{14}}
\renewcommand\large{\@setfontsize\large{10pt}{12}}
\renewcommand\footnotesize{\@setfontsize\footnotesize{7pt}{10}}
\makeatother

\renewcommand{\thefootnote}{\fnsymbol{footnote}}
\renewcommand\footnoterule{\vspace*{1pt}%
\color{cream}\hrule width 3.5in height 0.4pt \color{black}\vspace*{5pt}} 
\setcounter{secnumdepth}{5}

\makeatletter 
\renewcommand\@biblabel[1]{#1}            
\renewcommand\@makefntext[1]%
{\noindent\makebox[0pt][r]{\@thefnmark\,}#1}
\makeatother 
\renewcommand{\figurename}{\small{Fig.}~}
\sectionfont{\sffamily\Large}
\subsectionfont{\normalsize}
\subsubsectionfont{\bf}
\setstretch{1.125} 
\setlength{\skip\footins}{0.8cm}
\setlength{\footnotesep}{0.25cm}
\setlength{\jot}{10pt}
\titlespacing*{\section}{0pt}{4pt}{4pt}
\titlespacing*{\subsection}{0pt}{15pt}{1pt}

\fancyfoot{}
\fancyfoot[LO,RE]{\vspace{-7.1pt}\includegraphics[height=9pt]{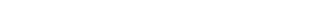}}
\fancyfoot[CO]{\vspace{-7.1pt}\hspace{13.2cm}\includegraphics{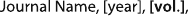}}
\fancyfoot[CE]{\vspace{-7.2pt}\hspace{-14.2cm}\includegraphics{head_foot/RF}}
\fancyfoot[RO]{\footnotesize{\sffamily{1--\pageref{LastPage} ~\textbar  \hspace{2pt}\thepage}}}
\fancyfoot[LE]{\footnotesize{\sffamily{\thepage~\textbar\hspace{3.45cm} 1--\pageref{LastPage}}}}
\fancyhead{}
\renewcommand{\headrulewidth}{0pt} 
\renewcommand{\footrulewidth}{0pt}
\setlength{\arrayrulewidth}{1pt}
\setlength{\columnsep}{6.5mm}
\setlength\bibsep{1pt}

\makeatletter 
\newlength{\figrulesep} 
\setlength{\figrulesep}{0.5\textfloatsep} 

\newcommand{\topfigrule}{\vspace*{-1pt}%
\noindent{\color{cream}\rule[-\figrulesep]{\columnwidth}{1.5pt}} }

\newcommand{\botfigrule}{\vspace*{-2pt}%
\noindent{\color{cream}\rule[\figrulesep]{\columnwidth}{1.5pt}} }

\newcommand{\dblfigrule}{\vspace*{-1pt}%
\noindent{\color{cream}\rule[-\figrulesep]{\textwidth}{1.5pt}} }

\makeatother

\twocolumn[
  \begin{@twocolumnfalse}
{\includegraphics[height=30pt]{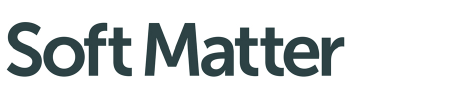}\hfill\raisebox{0pt}[0pt][0pt]{\includegraphics[height=55pt]{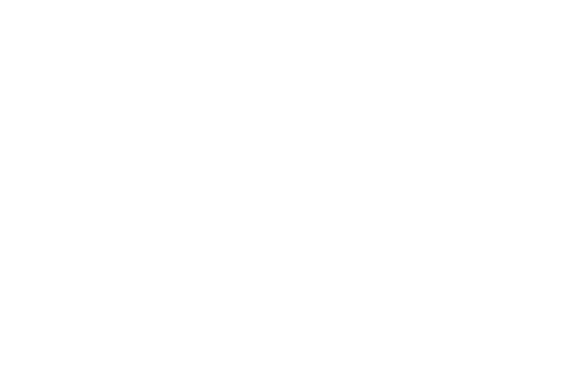}}\\[1ex]
\includegraphics[width=18.5cm]{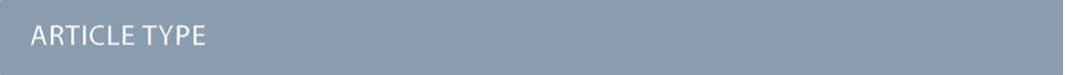}}\par
\vspace{1em}
\sffamily
\begin{tabular}{m{4.5cm} p{13.5cm} }

\includegraphics{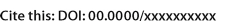} & \noindent\LARGE{\textbf{Dynamic Phases and Combing Effects for Elongated Particles Moving Over Quenched Disorder}} \\
\vspace{0.3cm} & \vspace{0.3cm} \\

 & \noindent\large{A. Lib{\' a}l,$^{\ast}$\textit{$^{a}$} S. Stepanov,\textit{$^{b}$} C. Reichhardt,\textit{$^{c}$} and C. J. O. Reichhardt$^{\ast}$\textit{$^{c}$}} \\

\includegraphics{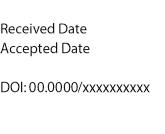} & \noindent\normalsize{We consider a two-dimensional system of elongated particles driven over a random quenched disorder landscape. For varied pinning site density, external drive magnitude, and particle elongation, we find a wide variety of dynamic phases, including random structures, stripe or combed phases with nematic order, and clogged states. The different regimes can be identified by examining nematic ordering, cluster size, number of pinned particles, and transverse diffusion. In some regimes we find that the pinning can enhance the particle alignment, producing a nonmonotonic signature in the nematic ordering with a maximum at a particular combination of pinning density and drive.  The optimal nematic occurs when a sufficient number of particles can be pinned, generating a local shear and leading to what we call a combing effect. At high drives, the combing effect is reduced when the number of pinned particles decreases. For stronger pinning, the particles form a heterogeneous clustered or clogged state that depins into a fluctuating state with high diffusion.} \\

\end{tabular}

 \end{@twocolumnfalse} \vspace{0.6cm}

  ]

\renewcommand*\rmdefault{bch}\normalfont\upshape
\rmfamily
\section*{}
\vspace{-1cm}


\footnotetext{\textit{$^{a}$~Mathematics and Computer Science Department, Babe{\c s}-Bolyai University, Cluj 400084, Romania}}
\footnotetext{\textit{$^{b}$~Physics Department, Babe{\c s}-Bolyai University, Cluj 400084, Romania}}
\footnotetext{\textit{$^{c}$~Theoretical Division and Center for Nonlinear Studies, Los Alamos National Laboratory, Los Alamos, New Mexico 87545, USA. Tel: 1 505 665 1134; E-mail: cjrx@lanl.gov. }}




A wide variety of systems can be modeled effectively
as an assembly of interacting particles driven over
random quenched disorder \cite{Fisher98,Reichhardt17}.
Specific examples include vortices in
type II superconductors \cite{Bhattacharya93,Koshelev94,Giamarchi96,Pardo98,Olson98a,Balents98},
Wigner crystals \cite{Reichhardt01}, skyrmions \cite{Reichhardt15a},
colloids \cite{Reichhardt02,Pertsinidis08,Bohlein12a,Tierno12},
pattern forming systems \cite{Reichhardt03,Zhao13},
sliding friction \cite{Vanossi13},
and active matter \cite{Bricard13,Sandor17a}. For systems with
long or intermediate range interactions, such as
superconducting vortices
and charged colloids,
a pinned phase appears for
drives less than a critical depinning force $F_{c}$, and
at higher drives disordered plastic
flow can occur with a combination of moving and pinned particles.
If the quenched disorder is weak, the depinning is elastic
and the particles keep their same neighbors at depinning
\cite{Fisher98,Reichhardt17,Fily10,DiScala12}.
In systems with plastic depinning,
there can be additional dynamic flow states
at higher drives when a partial dynamical ordering
occurs into a moving smectic or anisotropic crystal,
as observed
for driven
superconducting vortices
\cite{Koshelev94,Giamarchi96,Pardo98,Olson98a,Balents98},
skyrmions \cite{Reichhardt15a} and Wigner crystals \cite{Reichhardt01}.
When the substrate is periodic,
there can be multi-step depinning
transitions as well as
ordered and partially ordered moving states that can
take the form of one-dimensional or two-dimensional patterns
\cite{Reichhardt97,Gutierrez09,Bohlein12a}.
For
driven particles with short range or contact interactions,
such as non-charged colloids, granular
matter \cite{Yang17,McDermott19,McDermott20},
or active disks \cite{Sandor17},
dynamic transitions can occur
from a jammed or pinned phase at low drives to flowing 
density-modulated states,
where stripes can appear that are aligned with the direction of the drive.

In most studies for driven particles over quenched
disorder, the particle-particle interactions are
isotropic; however, there are also numerous
examples of systems where the particles
have anisotropic interactions or are elongated, such as the
rods or ellipses
found in different kinds of granular matter
\cite{Narayan07,Deseigne10,Kudrolli08,Borzsonyi13,Borzsonyi12,Nagy17,To21},
colloidal systems
\cite{Lowen94a,Sacanna11,Zheng11,Cohen11,Chen21},
and active matter systems \cite{Paxton04,Kumar14,Bar20,Arora22}.
Anisotropic interactions can also
arise in systems with longer range interactions
such as superconducting vortex liquid crystals
\cite{Carlson03,Reichhardt06a,Roe22},
electron liquid crystals \cite{Kivelson98,Lilly99a}, certain types of
magnetic skyrmion systems \cite{Lin15,Nagase19}, and
magnetic colloids in tilted fields \cite{Eisenmann04,Froltsov05}.
Despite the number of rod-like particles
or assemblies with one-dimensional
anisotropy that have been realized, there are almost no
studies of what happens when such
systems are driven over quenched disorder.

In this work, we consider a two-dimensional system of
elongated particles
modeled as five to nine connected disks that also
interact with $N_p$ randomly placed pinning sites.
We apply an increasing driving force and measure the number of
pinned particles, the number of particles in the largest
cluster, the nematic alignment, and the transverse diffusion.
We find a wide variety of dynamic phases, including
a random phase at low pinning densities and several varieties
of what we call combed phases at intermediate densities, where a
portion of the particles are pinned and produce local shearing of the
mobile particles, resulting in the emergence of alignment with the
driving direction.
At higher drives where all the particles are moving,
both the combing effect and the smectic ordering are reduced.
For high pinning densities, a clogged or arrested phase appears
in which the particle density is heterogeneous. This clogged state
depins into a disordered phase with intermediate nematic order.
Another interesting effect we observe is that
the nematic ordering is strongly nonmonotonic
as a function of pinning density and drive,
with random flow at low pinning density, maximal nematic ordering
for intermediate pinning densities and drives, and the reemergence of
disorder at high pinning densities.

\begin{figure}
  \centering
\includegraphics[width=3.5in]{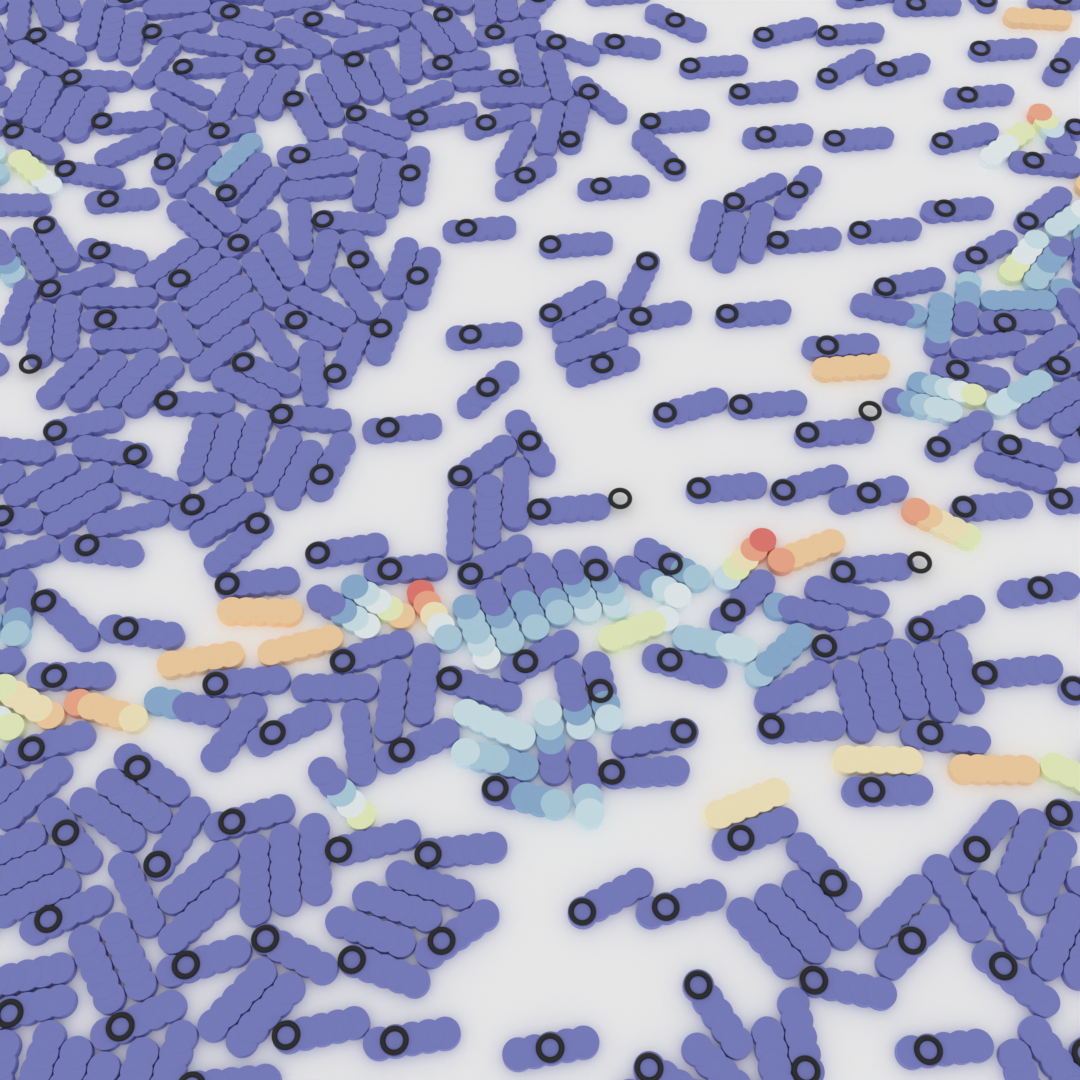}
\caption{
Image of a portion of the simulated system. The particles (colored) are constructed
out of overlapping two-dimensional disks assembled into a rigid object.
Disks belonging to one particle interact repulsively with disks on an adjacent
particle, generating a net force and torque on the center of mass of that
particle. 
Particles composed of $n=5$ discs are shown here.
A uniform drive $F_D$ along the $+x$ direction (to the right) causes the
particles to move over a substrate containing short range attractive
pinning sites (black circles).
Disk
color indicates instantaneous motion, with blue disks stationary and red disks
having the largest velocity. Particles with a range of colors across their
length are being subjected to torques.
}
\label{fig:1}
\end{figure}

\section{Simulation}
We consider a two-dimensional system of size $S_x=160$ and $S_y=160$
with periodic boundary conditions in the $x$ and $y$ directions.
This system, illustrated in Fig.~\ref{fig:1}, contains $N =400$ to $1200$
rigid elongated particles, each of
which is composed of
$n\in[5,7,9]$ overlapping circular disks of
radius $R_d = 1.0$ to give an aspect ratio of
$3:1$, $4:1$, and $5:1$, respectively. 
The disk locations for particle $i$ are calculated
according to
${\bf r}_{i,\alpha}={\bf R}_i+
(\alpha-(n-1)/2)R_d(\cos{\theta_i}{\bf \hat{x}}+\sin{\theta_i}{\bf \hat{y}})$,
where $\alpha=0 \ldots n-1$, ${\bf R}_i$ is the
center of mass of particle $i$, and $\theta_i$ is the angle
of particle $i$ with respect to the positive $x$ direction.

To obtain the interaction between two particles $i$ and $j$,
we calculate the pairwise repulsion ${\bf f}_{dd}^{(i,\alpha;j,\beta)}$ where
$\alpha$ ranges over all disks belonging to particle $i$ and $\beta$ ranges
over all disks belonging to particle $j$. These interactions are given by
a short-range stiff harmonic spring, so we obtain
${\bf{f}}_{dd}^{(i,\alpha)} = \sum_j^{N}k({\bf{R}}_{(i,\alpha;j,\beta)}-2R_d) \Theta(2R_d-R_{(i,\alpha;j,\beta)}){\bf \hat{R}}_{(i,\alpha;j,\beta)}$
where ${\bf{R}}_{(i,\alpha;j,\beta)}={\bf r}_{i,\alpha}-{\bf r}_{j,\beta}$,
$R_{(i,\alpha;j,\beta)}=|{\bf R}_{(i,\alpha;j,\beta)}|$,
${\bf \hat{R}}_{(i,\alpha;j,\beta)}={\bf R}_{(i,\alpha;j,\beta)}/R_{(i,\alpha;j,\beta)}$,
$\Theta$ is the Heaviside step function, and the elastic constant $k=10.0$.
Each of the constituent disks also interacts with 
a substrate modeled as $N_p$
randomly placed truncated parabolic attractive sites
of range $R_p = 0.5$ and maximum strength $F_p=1.0$:
${\bf f}_{\rm pin}^{(i,\alpha)}=-\sum_k^{N_p}(F_p/R_p){\bf R}_{(i,\alpha;k)}\Theta(R_p-R_{(i,\alpha;k)}){\bf \hat R}_{(i,\alpha;k)}$,
where ${\bf{R}}_{(i,\alpha;k)}={\bf r}_{i,\alpha}-{\bf r}_{k}^{(p)}$,
$R_{(i,\alpha;k)}=|{\bf R}_{(i,\alpha;k)}|$, and
${\bf \hat{R}}_{(i,\alpha;k)}={\bf R}_{(i,\alpha;k)}/R_{(i,\alpha;k)}$.
The radius of the pinning site is chosen to be small enough to prevent
multiple disks from being trapped by a single pinning site.
Each disk is also subjected to an external driving force
${\bf f}_{\rm ext}=F_D{\bf \hat x}$.

To update the position and orientation of a particle, we determine the
forces and torques exerted on the center of mass by the constituent
disks,

\begin{eqnarray}
  {\bf f}_i^{\alpha} & = & {\bf f}_{dd}^{(i,\alpha)}
    + {\bf f}_{\rm pin}^{(i,\alpha)} + {\bf f}_{\rm ext}\\
  {\bf f}_i & = & \sum_\alpha^n {\bf f}_i^{\alpha} \\
\tau_i & = & \sum_\alpha^n {\bf r}_i^\alpha \times {\bf f}_i^\alpha  
\end{eqnarray}

Next we update the center of mass and angular orientation
of the particle according to:
\begin{eqnarray} 
	{\bf R}_i(t+1) &= {\bf R}_i(t) + {\bf f}_i \Delta t/\eta\\
	\theta_i(t+1)  &= \theta(t) + \tau \Delta t/\eta
\end{eqnarray}

where we take the damping coefficient $\eta=1$ and $\Delta t=0.001$.
Finally we recalculate the positions
${\bf r}_{i,\alpha}$ of the constituent disks using the new
values of ${\bf R}_i$ and $\theta_i$, so that the elongated particles remain
completely rigid.

We initialize the system by placing
randomly oriented elongated particles at randomly chosen locations subject
to the constraint that there is no overlap between
constituent disks belonging to two different particles.
Our desired disk density is $\phi=0.4$, but the constraint method
gives a maximum possible density below this value. Thus,
in order to prepare denser samples,
we start from a lower density constrained state, allow it to evolve for
a period of time under the equations of motion described above, and then
insert additional particles into the free spaces created by the tendency of
the particles to cluster.
For example, in the case of $n=5$, we initialize the sample with $N=600$
particles and then introduce an additional 30 particles 
after each $t=100000$ simulation time steps
until we reach the desired density $\phi \simeq 0.4$
with $N = 1200$.
To obtain the same density in all systems,
for $n=7$ we use a total of $N=900$ particles,
and for $n=9$ we use $N=760$.
We do not perform any measurements until after all 
particles are present in the sample
and have been given a chance to move and compact.
The number of pinning sites in the sample ranges from $N_p=0$ to 600,
and the external drive ranges from
$f_d=0.01$ to 0.5.

\section{Results}
\begin{figure}
\centering
\includegraphics[width=3.5in]{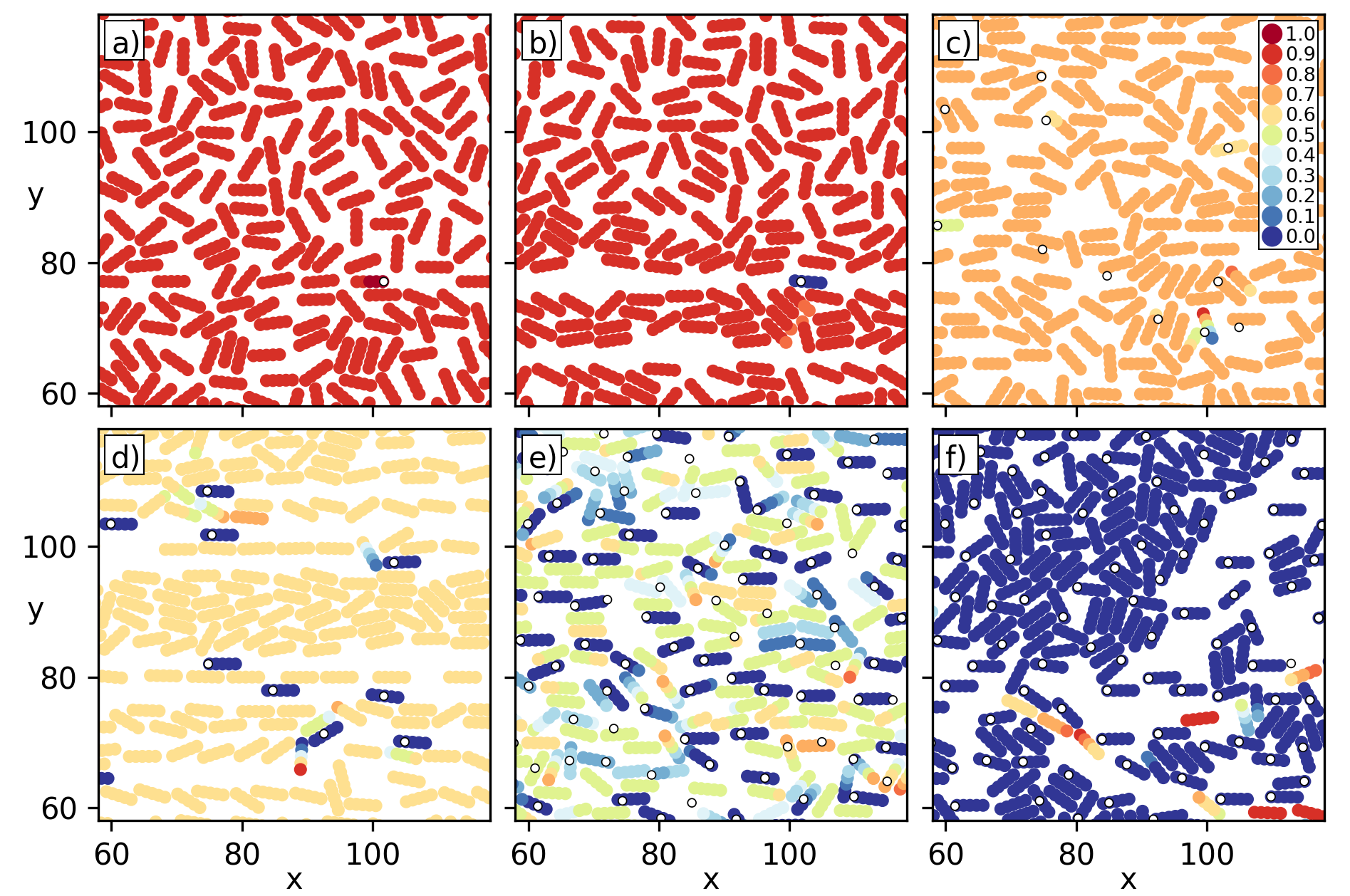}
\caption{
Simulation images illustrating the dynamic phases in an $n=5$ system.
The color of each disk indicates the distance moved by that disk over
a fixed period of time; red corresponds to ballistic motion in the direction
of drive, and blue corresponds to stationary.
For clarity, only a $60 \times 60$ subsection of the full
$160 \times 160$ system is shown.
(a) Phase I (random ballistic) for $N_p=3$ and $F_{D} = 0.25$.
(b) Phase II (locally combed ballistic) for $N_p=3$ and $F_{D} = 0.03$.
(c) Phase III (point combed)
for $N_p=50$ and $F_{D} = 0.35$.
(d) Phase IV  (tooth combed) for $N_p=30$ and $F_{D} = 0.07$.
(e) Phase VI (channel combed)
for $N_p=600$ and $F_{D} = 0.12$.
(f) Phase VII (clogged) for
$N_p=600$ and $F_{D} = 0.02$.
}
\label{fig:2}
\end{figure}

In Fig.~\ref{fig:2}
we highlight the six dynamical phases
we find for $n=5$.
For a low pinning density of $N_p=3$, at $F_D=0.25$
the particles form
a random structure that moves ballistically, which we call Phase I or the random ballistic phase, illustrated in
Fig.~\ref{fig:2}(a).
At $F_D=0.3$, Fig.~\ref{fig:2}(b) shows that the sample has entered Phase II
where 
the pinning sites each permanently capture one disk at the end of a particle,
and
the driving force causes the pinned particles to align in the
direction of the drive.
The small number of pinning sites
create a low-density stripe state
via a local combing effect
in which there is a tendency for freely moving particles that collide with
pinned particles
to become aligned with the drive.
We label this Phase II, the locally combed ballistic phase,
in which most of the particles are moving
at the driving velocity.
In the point combed state or Phase III, shown in
Fig.~\ref{fig:2}(c) for $N_p = 50$ and $F_{D} = 0.35$,
disks are no longer permanently trapped by the pinning sites but the
pinning interactions
slow the motion of the particles.
Here the average particle velocity
drops below the ballistic limit due to the
pinning site collisions,
and there is some weak nematic ordering due to the combing effect.

Figure~\ref{fig:2}(d) shows Phase IV
or the tooth combed state at $N_{p} = 30$ and $F_{D} = 0.07$.
There are now some permanently pinned particles
and the pinning density is large enough to induce strong
nematic ordering.
In the combed channel flow phase VI, illustrated
in Fig.~\ref{fig:2}(e) at $N_p = 600$ and $F_{D} = 0.12$, 
a number of particles are pinned and there is some local nematic ordering,
but the flow occurs plastically through channels.
Figure~\ref{fig:2}(f) shows phase VII or the clogged state
at $N_{p} = 600$ and $F_{D} = 0.02$,
where now most of the particles are permanently pinned
in local clusters, giving a heterogeneous state composed of coexisting
regions of high density and low density,
similar to the clogged states
observed for monodisperse individual disks
moving through random obstacle arrays
\cite{Peter18}.
For high $N_p$ and high $F_D$,
all of the particles are moving
and we observe phase V (not shown)
where the particle positions are
uniformly random, similar to phase I,
but due to the collisions with the pinning sites,
the particles gradually diffuse with respect to each other.
For phase I at low $N_p$, there is almost no diffusion. 

\begin{figure}
\centering
\includegraphics[width=3.5in]{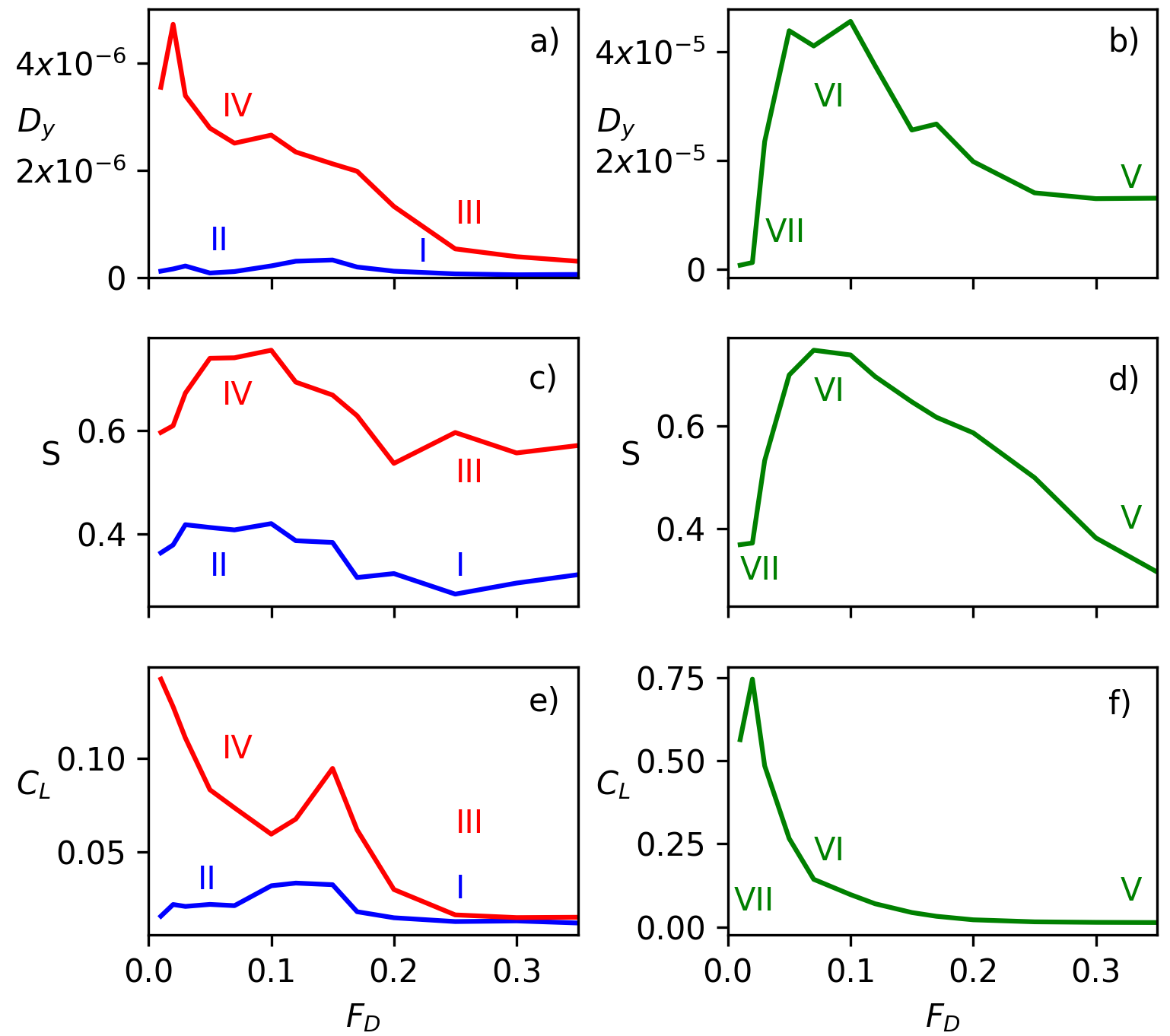}
\caption{
Measures of the response of particles of size $n=5$ as a function of $F_D$
showing the locations of phases I, II, III, IV, V, VI, and VII.
(a) Transverse diffusion $D_y$ vs $f_{d}$ for $N_p = 5$ (blue) and $50$ (red).
(b) $D_{y}$ vs $F_{D}$ for $N_p = 600$.
(c) The particle alignment $S$ vs $F_{D}$ for $N_p = 5$ (blue) and $50$ (red).
(d) $S$ vs $F_{D}$ for $N_p = 600$.
(e) Fraction of particles in the largest cluster $C_{L}$ vs $F_{D}$
for $N_p = 5$ (blue) and $50$ (red).
(f) $C_{L}$ vs $F_{D}$ for $N_{p} = 600$.
}
\label{fig:3}
\end{figure}

We can characterize
the different states by
measuring $D_y$, the mean square displacement in the direction
perpendicular to the drive; $S$,
the alignment of the particles obtained from
the nematic order parameter $P_2(\cos(\theta)) = 1 - \cos^2(\theta)$;
and $C_L$, the fraction of particles in the largest
contiguous cluster.
In Fig.~\ref{fig:3}(a,b)
we plot $D_{y}$ versus $F_{D}$ for the
$n=5$ sample from Fig.~\ref{fig:2}
for $N_p = 5$, $50$, and $600$.
Figure~\ref{fig:3}(c,d) shows
the corresponding $S$ versus $F_{D}$ measures, while in
Fig.~\ref{fig:3}(e,f)
we plot $C_{L}$ versus $F_{D}$.
We also highlight the locations of phases I through VII.
For $N_{p} = 5$, 
$D_{y}$, $S$, and $C_{L}$ are low,
only phases II and I occur,
and there is a small decrease of
$S$ in phase I.
There is a small number of pinned particles
in phase II that nucleate localized nematic ordering,
and the drop in $S$
upon entering phase I occurs when all the particles begin to move and
the local combing effect is lost.
For $N_p = 50$, $D_y$ decreases with
increasing $F_{D}$ and $S$ passes through a maximum near $F_{D}= 0.11$.
Phase IV is associated with a large value of
$S$ and intermediate values of $C_L$ and $D_y$, while
in phase III where all of the particles are moving,
$C_L$ and $D_y$ are small.
$S$ is lower in phase III than 
in phase IV due to the reduced combing effect.
There is a small
feature in $C_L$ near $F_{D} = 0.15$ at the transition between
phases IV and III.
For $N_p = 600$,
$D_{y}$ and $S$ are both low in the clogged phase VII,
but $C_{L}$ is large since the system
forms a large pinned cluster.
As the drive increases, the particles
depin and $C_{L}$ decreases,
while $S$ and $D_y$ pass through a local peak in phase VI
and then decrease as the drive
is further increased. At the higher drives,
all the particles are moving and $S$ and $C_{L}$
are low,
but $D_y$ remains
finite in phase V due to the collisions with the pinning sites.

\begin{figure}
\centering
\includegraphics[width=3.5in]{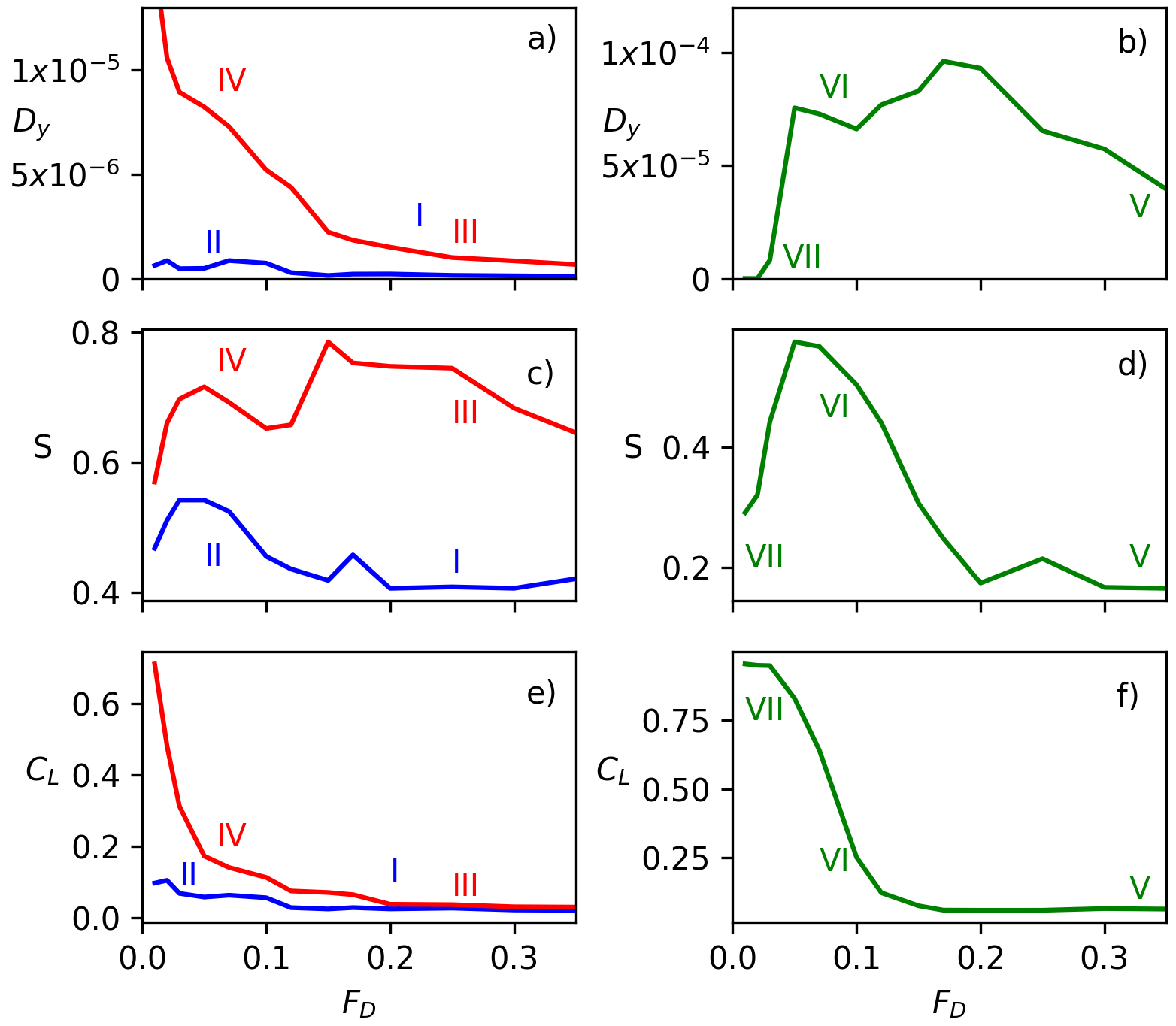}
\caption{
Measures of the response of particles of size $n=9$ as a function of
$f_d$
showing the locations of phases I, II, III, IV, V, VI, and VII.
(a) $D_y$ vs $F_{D}$ for $N_p = 5$ (blue) and $50$ (red).
(b) $D_{y}$ vs $F_{D}$ for $N_p = 600$.
(c) $S$ vs $F_{D}$ for $N_p = 5$ (blue) and $50$ (red).
(d) $S$ vs $F_{D}$ for $N_p = 600$.
(e) $C_{L}$ vs $F_{D}$ for $N_p = 5$ (blue) and $50$ (red).
(f) $C_{L}$ vs $F_{D}$ for $N_{p} = 600$.
}
\label{fig:4}
\end{figure}

In Fig.~\ref{fig:4} we plot $D_y$, $S$, and $C_L$ versus $F_D$
for particles with $n=9$ moving over landscapes with
$N_p=5$, 50, and 600, where we observe similar trends as in the
$n=5$ system.
Here, $S$ shows a strong drop for $N_p=600$ at
$F_D=0.35$ where a transition occurs from
phase VI to phase V,
while at the same drive for $N_p=50$,
$S$ is still large and the system is in phase III.
The diffusion $D_y$ in phase $V$
for $N_p=600$
is high at $F_{D} = 0.35$ but $S$ and $C_L$ are small, while
for $N_p = 5$ and $N_p=50$, the diffusion is low at the same drive.
For particles of length $n=7$, we also observe similar behavior (not
shown).

\begin{figure}
\centering
\includegraphics[width=3.5in]{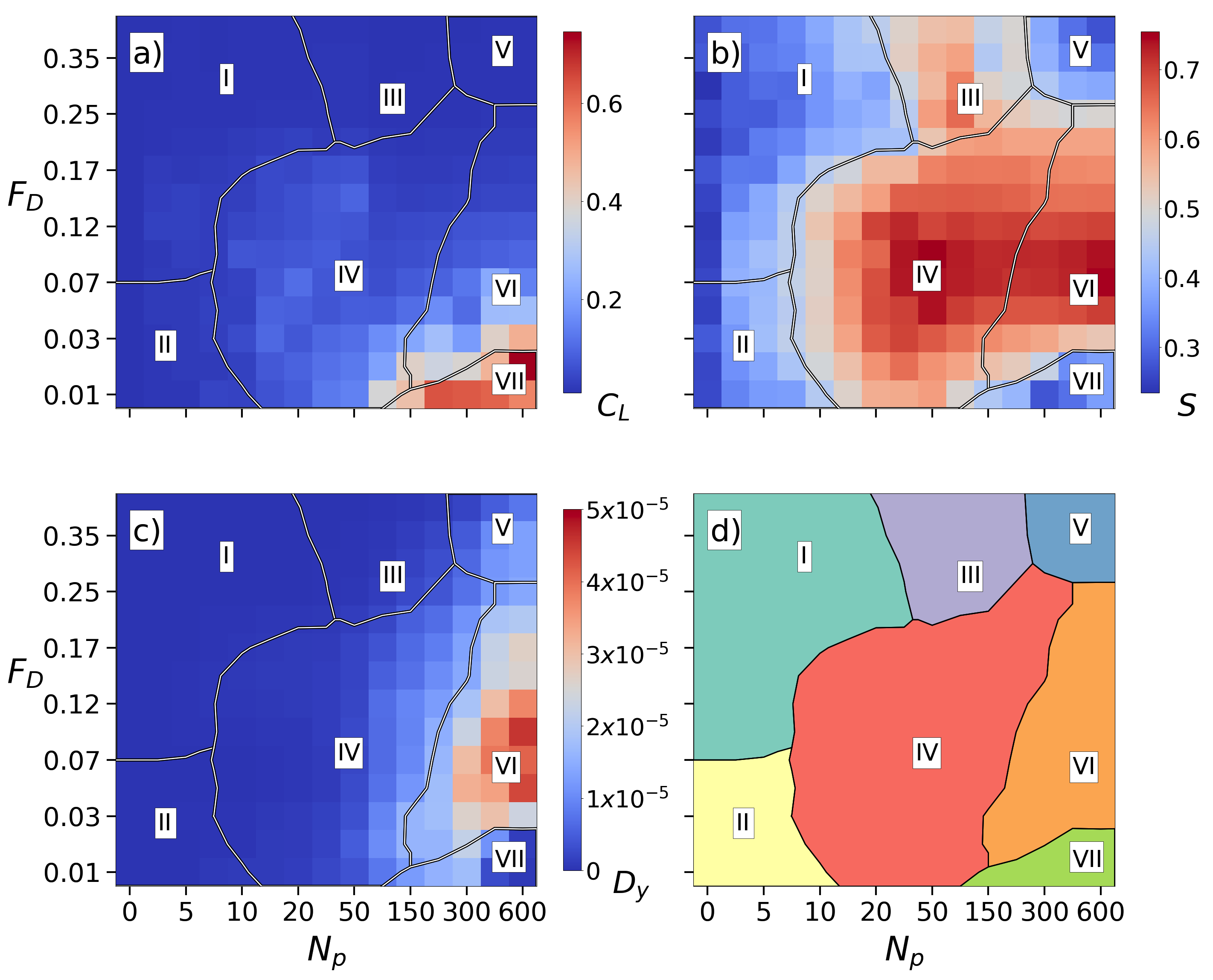}
\caption{
(a,b,c) Heat maps as a function of $F_D$ vs $N_p$ for the
$n=5$ system from Fig.~\ref{fig:2}.  
(a) $C_L$. 
(b) $S$.
(c) $D_y$.
(d) Schematic phase diagram constructed from the above measures showing
the locations of phases
I (random ballistic), II (locally combed ballistic),
III (point combed), IV (tooth combed), V (uniform random),
VI (combed channel), and VII (clogged).
}
\label{fig:5}
\end{figure}

From measurements of $D_y$, $S$, and $C_{L}$ versus $F_{D}$,
we can construct dynamic phase diagrams for the different particle lengths.
Fig.~\ref{fig:5}(a,b,c) shows heat diagrams of the
cluster size $C_L$, alignment $S$, and traverse diffusion $D_y$
as a function of $F_D$ versus $N_p$ for
the $n=5$ system from Fig.~\ref{fig:2}.
The resulting schematic phase diagram in Fig.~\ref{fig:5}(d)
highlights the regions where phases I through VII appear as a function
of $F_D$ versus $N_p$.
When $N_p<40$,
phase I occurs for $F_D>=0.7$,
where $C_L$ and $S$ are small and there is no diffusion,
while phase II appears for $F_{D} < 0.07$, where some of the particles
are permanently pinned and there is weak
alignment.
In the clogged phase VII, which exists in the range
$N_p>200$ and $f_D<0.03$, $C_L$ is high while
$S$ and $D_y$ remain small.
The tooth combed phase IV for intermediate $N_p$ and
$F_D<0.16$,
where some of the particles can become pinned temporarily,
has high $S$.
Figure~\ref{fig:5}(b) illustrates that $S$ is nonmonotonic as a
function of both $F_D$ and $N_p$,
with the highest value occurring in phase IV.
In phase III, $S$ is still large but there are no pinned
particles present.
The combed channel phase VI occurs for $0.3 < F_{D} < 0.15$, above the
depinning transition from the clogged state, and has
high $D_y$, low $C_L$, and high $S$.
Phase V is where the drive is strong enough that all the particles are moving 
and the alignment is lost; however, there is still some weak diffusion 
due to the collisions with the pinning sites. 

\begin{figure}
\centering  
\includegraphics[width=3.5in]{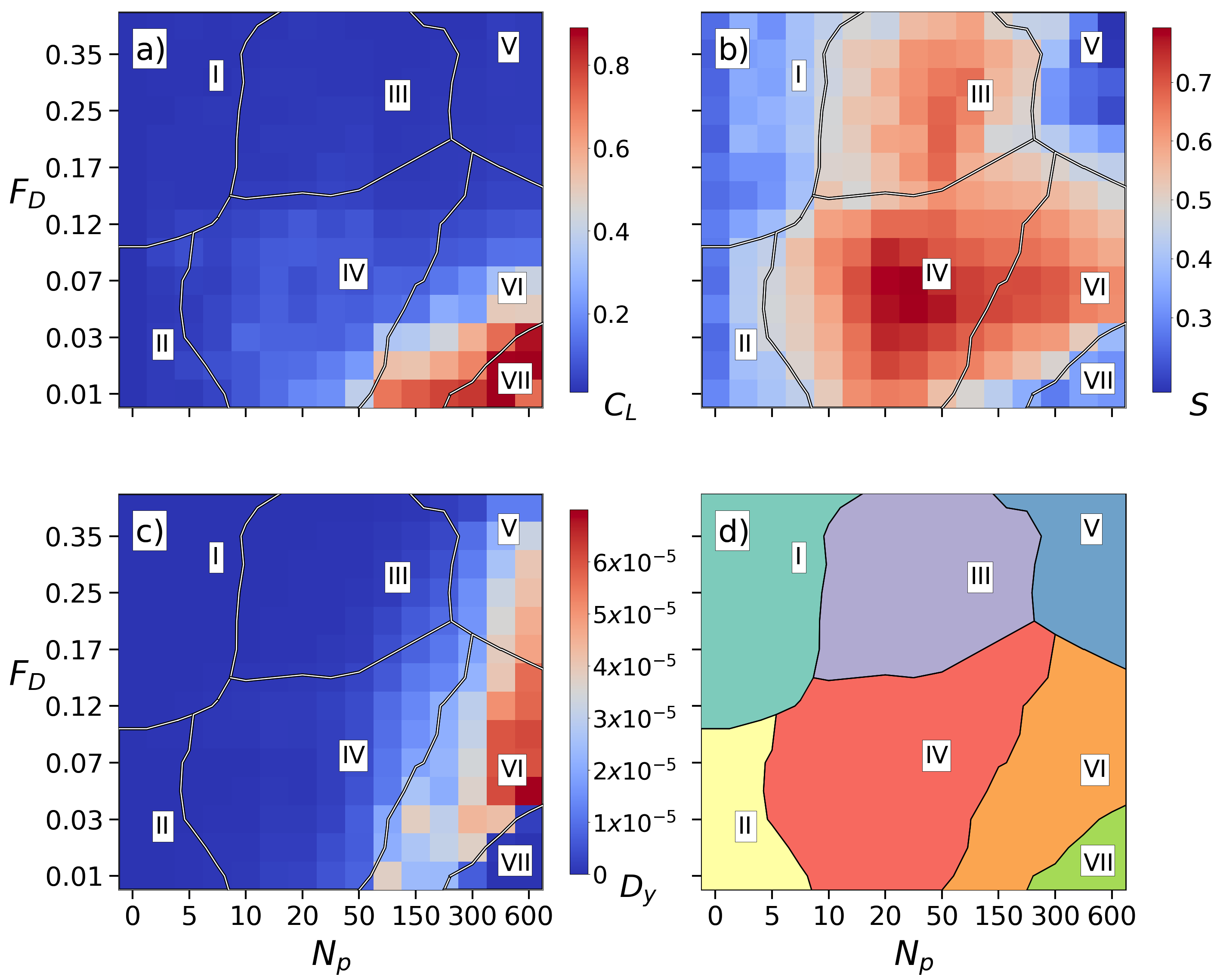}
\caption{
(a,b,c) Heat maps as a function of $F_D$ vs $N_p$ for the $n=7$
system.
(a) $C_L$. (b) $S$. (c) $D_y$.  
(d) Schematic phase diagram constructed from the above measures
showing the locations of phases I (random ballistic), II (locally combed
ballistic), III (point combed), IV (tooth combed), V (uniform random),
VI (combed channel), and VII (clogged).
}
\label{fig:6}
\end{figure}

\begin{figure}
\centering  
\includegraphics[width=3.5in]{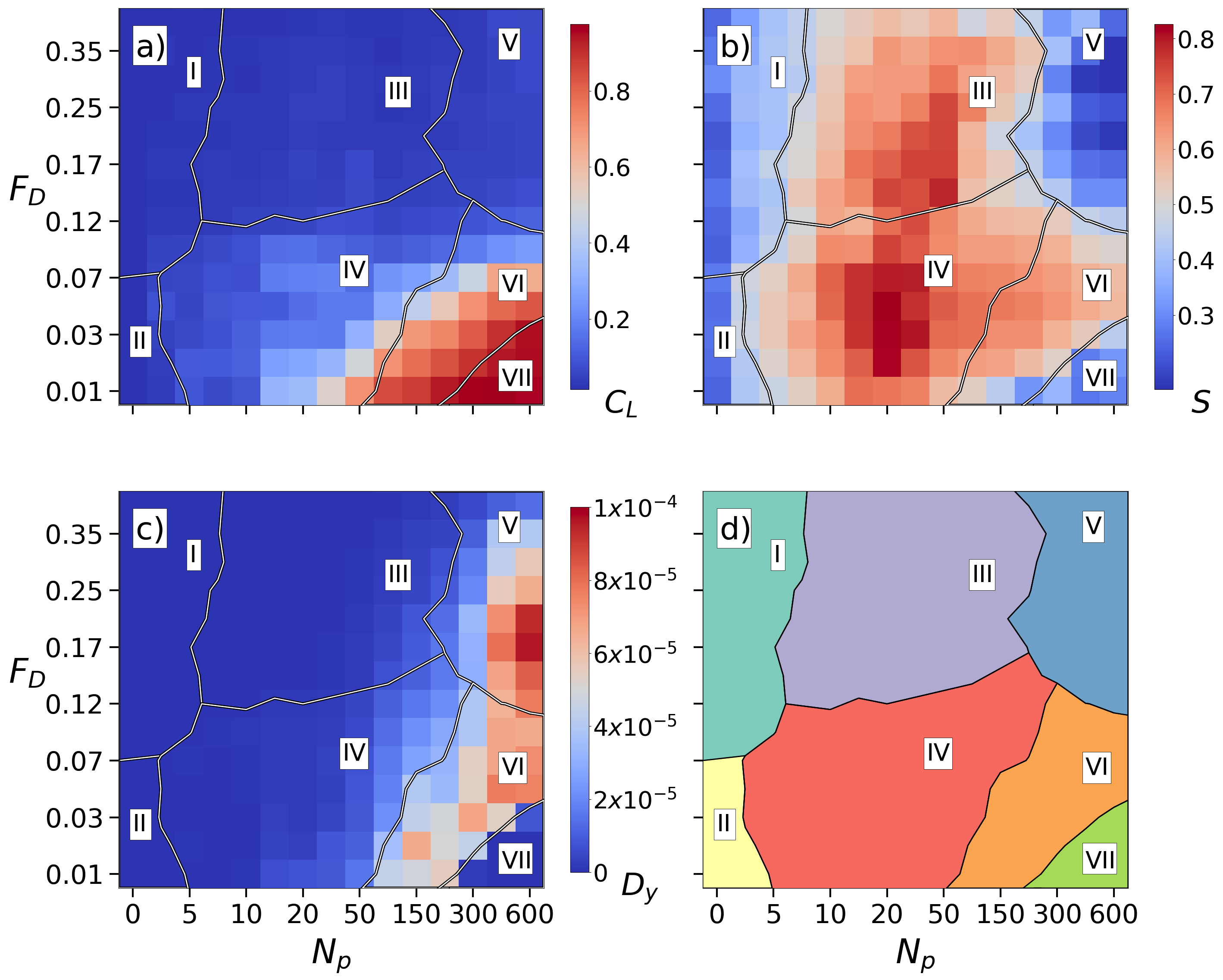}
\caption{
Heat maps as a function of $F_D$ vs $N_p$ for the $n=9$ system.
(a) $C_L$. (b) $S$. (c) $D_y$. (d) Schematic
phase diagram constructed from the above measures showing the locations
of phases I (random ballistic), II (locally combed ballistic),
III (point combed), IV (tooth combed), V (uniform random), VI (combed
channel), and VII (clogged).
}
\label{fig:7}
\end{figure}

In Fig.~\ref{fig:6}, the heat maps of $C_L$, $S$, and $D_y$ along with the
phase diagram as a function of $F_D$ versus $N_p$
for a sample with $n=7$ show the same general trends found in
Fig.~\ref{fig:5} for the $n=5$ particles.
Figure~\ref{fig:7} is for the $n=9$ particles and shows the same
quantities of $C_L$, $S$, and $D_y$ plotted as heat maps as a function
of $F_D$ versus $N_p$ along with a schematic phase diagram.
The tooth combed phase IV becomes wider as $n$ increases since
the longer particles can be combed more effectively by the pinning
sites. The uniform random phase V also becomes
more extensive in size with increasing $n$ since the pointlike pinning sites
decouple at lower drives from the more elongated particles.
Overall, these results indicate that the generic phases we observe
remain robust for a range of particle lengths.

\section{Discussion}
The dynamic phase diagrams we find for the elongated particles
show several distinct differences from those
measured for isotropic particles.
Yang {\it et al.} \cite{Yang17} considered a  model of hard
disks driven over quenched disorder
for varied ratios of
the number of pinning sites to the number of disks and observed
a pinned state along with a plastic phase,
a density modulated phase, and a moving smectic
phase. In general,
the phase diagram for the hard disks was similar to that
found in systems with longer
range interactions such as
superconducting vortices and colloids \cite{Reichhardt17},
where at high drives, a dynamically reordered moving smectic
can form.
The most noticeable difference
between the isotropic short-range interacting disks and longer-range 
interacting particles is the density-modulated phases,
which are absent in systems with longer range repulsion since
for those systems the
energy is minimized when the system density remains fairly uniform.
In the case we consider here of elongated grains,
the most similar phase to the moving smectic is the
tooth combed phase IV, which shows
nonmonotonic behavior as a function of drive and
pinning number.
In contrast, the smectic phase
observed for the isotropic disks varies monotonically with both quantities.
We also note that for particle densities much higher than what we
consider here,
the optimal packing would be aligned rods;
in that case, the system could act like a
rigid solid, and the behavior would strongly depend
on how the system is initially prepared.
It could also be interesting to
consider mixtures of rods of different lengths or
monodisperse disks interspersed with rods,
where different regimes involving phase separation could arise.
We concentrated on varying the pinning density over a range of drives;
however, since the combed phases
occur when some particles are permanently pinned,
this suggests that some of the phases could
also be realized by introducing
obstacles or elongated obstacles instead of pinning
sites. It is likely that if obstacles were used, there
would be much more extended regions of clogged phase.

\section{Conclusions}
We have investigated the dynamics of elongated particles
modeled as connected hard disks and driven
over random pinning.
As a function of pinning number and drive, we
observe a variety of phases, including
a random phase at low pinning densities.
In the combed phases,
the pinning sites capture some particles that become aligned
with the direction of the driving force
and facilitate
the spreading of nematic alignment
to the surrounding flowing particles.
The nematic
ordering is strongly non-monotonic
as a function of pinning number and drive, and
reaches a maximum for drives that are small enough for some of
the particles to remain pinned and pinning numbers that
are small enough that the entire system does not become
disordered.
At high drives, the nematic ordering is reduced once all the particles
are in motion due to the
loss of the combing effect from the pinned particles.
We also observe a heterogeneous clogged phase,
a plastic moving phase, and a high drive random fluid-like phase
for large pining numbers.
The dynamic phases we find show several differences from
those known from isotropically interacting 
particles moving over random disorder,
including the non-monotonic nematicity.
Our results should be relevant to anisotropic colloids,
grains, and other driven systems with one-dimensional anisotropic
particle interactions. 




\section*{Conflicts of interest}
There are no conflicts to declare.

\section*{Acknowledgements}
This work was supported by the US Department of Energy through
the Los Alamos National Laboratory.  Los Alamos National Laboratory is
operated by Triad National Security, LLC, for the National Nuclear Security
Administration of the U. S. Department of Energy (Contract No. 892333218NCA000001).
AL was supported by a grant of the Romanian Ministry of Education
and Research, CNCS - UEFISCDI, project number
PN-III-P4-ID-PCE-2020-1301, within PNCDI III.



\balance


\bibliography{mybib} 
\bibliographystyle{rsc} 

\end{document}